\documentclass{myamc}

\usepackage[latin1]{inputenc}
\usepackage[T1]{fontenc}
\usepackage{ae,aecompl}
\usepackage{subfig}
\usepackage{booktabs}
\usepackage{amsmath}
\usepackage{paralist}
\usepackage{graphics}
\usepackage{epsfig}
\usepackage{hyperref}

\newtheorem{theorem}{Theorem}[section]

\theoremstyle{definition}
\newtheorem{definition}[theorem]{Definition}

\newtheorem{example}{Example}

\DeclareMathOperator{\GF}{GF}
\DeclareMathOperator{\wt}{wt}
\DeclareMathOperator{\Aut}{Aut}
\DeclareMathOperator{\tr}{tr}
\DeclareMathOperator{\symp}{Sp}
\newcommand{\ket}[1]{\left| #1 \right>}
\newlength{\skpl}
\newcommand{\skp}{\hspace{\skpl}}

\title[Classification of self-dual additive codes]{Graph-based classification of self-dual additive codes over finite fields}
\author[Lars Eirik Danielsen]{}
\subjclass{Primary: 94B05, 05C30; Secondary: 94B60, 05C50}
\keywords{Additive codes, self-dual codes, quantum codes, stabilizer states, graphs, local complementation, circulant codes, MDS codes}
\thanks{The author is supported by the Research Council of Norway.}

\begin{document}

\maketitle

\centerline{\scshape Lars Eirik Danielsen}
\medskip
{\footnotesize
  \centerline{Department of Informatics}
  \centerline{University of Bergen}
  \centerline{PO Box 7803, N-5020 Bergen, Norway}
}

\bigskip

\centerline{}

\begin{abstract}
Quantum stabilizer states over $\mathbb{F}_m$ can be represented as self-dual additive codes over $\mathbb{F}_{m^2}$.
These codes can be represented as weighted graphs, and orbits of graphs under the generalized 
local complementation operation correspond to equivalence classes of codes. We have previously 
used this fact to classify self-dual additive codes over $\mathbb{F}_4$. In this paper we classify self-dual 
additive codes over $\mathbb{F}_9$, $\mathbb{F}_{16}$, and $\mathbb{F}_{25}$.
Assuming that the classical MDS conjecture holds, we are able to classify all self-dual additive MDS codes 
over $\mathbb{F}_9$ by using an extension technique. 
We prove that the minimum distance of a self-dual additive code is related to the minimum vertex
degree in the associated graph orbit. Circulant graph codes are introduced, 
and a computer search reveals that this set contains many strong codes.
We show that some of these codes have highly regular graph representations.
\end{abstract}

\section{Introduction}

It is well-known that \emph{self-orthogonal additive codes} over $\mathbb{F}_4$
can be used to represent a class of \emph{quantum error-correcting codes} known
as \emph{binary stabilizer codes}~\cite{calderbank}.
Although the binary stabilizer codes have been studied most, several authors have
considered nonbinary stabilizer codes over finite fields~\cite{nonbin3,grassloptimal,grasslpaper,nonbin2,nonbin1,
schlingemann2}, cyclic groups~\cite{hostens}, and
Abelian groups in general~\cite{schlingemann}. We will focus mainly on codes over finite fields, 
and exploit the fact that a stabilizer code over $\mathbb{F}_m$ corresponds to a self-orthogonal \emph{additive 
code} over $\mathbb{F}_{m^2}$. Quantum codes of dimension zero are known 
as \emph{stabilizer states}, which are entangled quantum states with several possible 
applications. Stabilizer states correspond to \emph{self-dual} additive codes. 
It is known that such codes can be represented as graphs~\cite{grasslpaper,schlingemann}. 
It is also known that two self-dual additive codes over $\mathbb{F}_4$ are equivalent
if and only if their corresponding graphs are equivalent, up to isomorphism, with respect to a sequence
of \emph{local complementations}~\cite{bouchet4,glynnbook,nestthesis,nest}.
We have previously used this fact to devise a graph-based algorithm with which we classified all self-dual
additive codes over $\mathbb{F}_4$ of length up to 12~\cite{selfdualgf4}.
Recently, the representation of equivalence classes as graph orbits was generalized to
self-dual additive codes over any finite field~\cite{newlc}.
In this paper we use graph-based algorithms to classify all self-dual
additive codes over $\mathbb{F}_9$, $\mathbb{F}_{16}$, and $\mathbb{F}_{25}$ up to lengths 8, 6, and 6, respectively.
We also give upper bounds on the number of codes, derived from \emph{mass formulas}.
By using a graph extension technique we find that there are only three non-trivial self-dual additive MDS codes 
over $\mathbb{F}_9$, assuming that the classical MDS conjecture holds.
We prove that the minimum distance of a self-dual additive code is related to the minimum vertex
degree in the associated graph orbit.
Finally, we perform a search of \emph{circulant} graph codes, 
a subclass of the self-dual additive codes, which is shown to contain many codes with high minimum distance. 
The highly regular graph structures of some of these codes are described.

\section{Stabilizer states}

Data in a classical computer are typically stored in bits that have values either 0 or~1.
Similarly, we can envisage a quantum computer where data are stored in quantum bits, also known as \emph{qubits}, 
i.e., two-level quantum systems. One qubit can then be described by a 
vector $\ket{x} = \binom{\alpha}{\beta} \in \mathbb{C}^2$, 
where $|\alpha|^2$ is the probability of observing the value 0 when we measure the qubit, 
and $|\beta|^2$ is the probability of observing the value 1. More generally, data could be 
stored in $m$-level \emph{qudits}, described by vectors from $\mathbb{C}^m$. Measuring such a 
qudit would give a result from an alphabet with $m$ symbols. In general, this alphabet
could be any finite Abelian group, but we will focus on the case where the alphabet is a finite field.
The $m$ vectors $\ket{x}$, $x \in \mathbb{F}_m$, form an orthonormal basis of $\mathbb{C}^m$.

An error operator that can affect a single qudit is represented by a complex \emph{unitary} $m \times m$ matrix, 
i.e., a matrix $U$ such that $U U^{\dag} = I$, where $\dag$ means conjugate transpose.
A state of $n$ qudits is represented by a vector
from $\mathbb{C}^{m^n} = \mathbb{C}^m \otimes \cdots \otimes \mathbb{C}^m$.
Assuming that errors act independently on each qubit, this state is
affected by error operators described by $n$-fold tensor products of unitary $m \times m$ matrices.
In the case of qubits ($m=2$), we only need to consider errors from the \emph{Pauli group},
\[
X=\begin{pmatrix} 0 & 1 \\ 
                  1 & 0 \end{pmatrix},\quad
Z=\begin{pmatrix}  1 & 0 \\ 
                   0 & -1\end{pmatrix},\quad
Y=iXZ=\begin{pmatrix}  0 & -i \\ 
                      i & 0\end{pmatrix},\quad
I=\begin{pmatrix}  1 & 0 \\ 
                   0 & 1\end{pmatrix},
\]
due to the fact that these matrices form a basis of all unitary $2 \times 2$ matrices.
The error $X$ is called a \emph{bit-flip}, since $X \ket{x} = \ket{x+1}$.
The error $Z$ is known as a \emph{phase-flip}, since $Z \ket{x} = (-1)^x \ket{x}$.
For general qudits that take their values from $\mathbb{F}_m$, we consider the \emph{generalized Pauli group}, 
$\mathcal{P}_m$, also known as the \emph{discrete Heisenberg-Weyl group}. 
When our alphabet is a finite field, we must have $m=p^r$, where $p$ is a prime and $r \ge 1$.
The errors contained in the generalized Pauli group are \emph{shift errors}, $X(a)\ket{x} = \ket{x+a}$,
and \emph{phase errors}, $Z(b)\ket{x} = \omega^{\tr_{m/p}(bx)} \ket{x}$, where $a,b \in \mathbb{F}_m$, 
$\omega$ is a complex $p$-th root of unity, and $\tr_{m/p}: \mathbb{F}_m \to \mathbb{F}_p$ is the trace 
function, $\tr_{m/p}(c) = \sum_{i=0}^{r-1} c^{p^i}$.
If $m=p$ is a prime, i.e., $r=1$, the generalized Pauli group is generated by
\[
\left<X(1)=\begin{pmatrix} 0 &   &        & \\ 
                      \vdots&   & I_{m-1} & \\
                          0 &   &        & \\
                          1 & 0 & \cdots & 0 \end{pmatrix},\quad
Z(1)=\begin{pmatrix}  1 &        &          &        & 0 \\ 
                       & \omega &          &        & \\
                       &        & \omega^2 &        & \\
                       &        &          & \ddots & \\
                     0 &        &          &        & \omega^{n-1}\end{pmatrix}\right>,
\]
where $\omega$ is a complex $p$-th root of unity, and $I$ is the identity matrix of specified dimension.\footnote{The 
set of generators also contains the scalar $\omega$,
except for the case $m=2$, where it contains $i$, a 4-th root of unity. 
This overall phase factor can be ignored for our purposes.} The operators $X(a)$ and $Z(b)$ 
are obtained by taking the $a$-th and $b$-th powers of $X(1)$ and $Z(1)$, respectively.
Even if $m$ is not prime, we can still define qudits that take values from the cyclic group $\mathbb{Z}_m$,
and use the same error operators as defined above. However, when $m$ is a prime power, 
we get much better codes by using a finite field as our alphabet.
When we work with qudits that take values from $\mathbb{F}_{p^r}$, where $r>1$, 
we use the error group $\{ \bigotimes_{i=1}^{r} E_i \mid E_i \in \mathcal{P}_p \}$~\cite{nonbin2},
i.e., the operators are $r$-fold tensor products of Pauli matrices from the group $\mathcal{P}_p$.
The error bases that we use are examples of \emph{nice error bases}~\cite{nicebases}.

\emph{Quantum codes} are designed to add redundancy in order to protect quantum states against errors 
due to interference from the environment. A code of \emph{length}~$n$ and \emph{dimension}~$k$
adds redundancy by encoding $k$ qudits using $n$ qudits. One type of code that exploits the fact 
that the generalized Pauli group forms a basis for all possible errors is the \emph{stabilizer code}~\cite{gottesman}.
A \emph{stabilizer} is an Abelian group generated by a set of $n-k$ commuting error operators.
An error is detected by measuring the eigenvalues of these operators. If a state is a valid codeword that
has not been affected by error, we will observe the eigenvalue $+1$ for all operators. The quantum code, i.e.,
the set of all valid codewords, is therefore a joint eigenspace of the stabilizer.
If there is a detectable error, some eigenvalues would be different from $+1$, due to the commutativity
properties of the generalized Pauli matrices.
A stabilizer generated by a set of $n$ error operators defines a zero-dimensional
quantum code, also known as a \emph{stabilizer state}.\footnote{Stabilizer states
could also be called one-dimensional quantum codes, since they are one-dimensional Hilbert subspaces. We use
the term dimension to mean the number of qudits the code can encode.}
The \emph{minimum distance} of a zero-dimensional stabilizer code is simply
the minimum nonzero \emph{weight} of all error operators in the stabilizer. The weight of an error
operator is the number of $m \times m$ tensor components that are different from the identity matrix.
A quantum code of length~$n$, dimension~$k$, and minimum distance~$d$, over the alphabet $\mathbb{F}_m$, is
denoted an $[[n,k,d]]_m$ code. Stabilizer states are therefore $[[n,0,d]]_m$ codes.
If the minimum distance, $d$, is high, the stabilizer state is robust against error, which
indicates that it is highly \emph{entangled}. Entangled quantum states have many potential applications, for instance
in cryptographic protocols, or as \emph{graph states}~\cite{cluster} which can be used as a 
resource for quantum computations. In the next section we will also see that zero-dimensional stabilizer codes 
correspond to an interesting class of classical codes, known as \emph{self-dual additive codes}.

\begin{example}\label{ex:stabilizer}
A $[[4,0,3]]_3$ stabilizer state is obtained from the stabilizer generated by the following error operators.
\begin{alignat*}{4}
& X(1) &&\otimes X(1)Z(2) &&\otimes I &&\otimes X(1), \\
& X(1)Z(1) &&\otimes X(2) &&\otimes X(1)Z(1) &&\otimes X(1), \\
& I &&\otimes X(2)Z(2) &&\otimes X(1)Z(1) &&\otimes Z(2), \\
& X(1) &&\otimes X(2)Z(2) &&\otimes X(2) &&\otimes X(2)Z(2).
\end{alignat*}
\end{example}

\section{Self-dual additive codes}

We can represent a stabilizer state over $\mathbb{F}_m$ by an $n \times 2n$ matrix $(A \mid B)$~\cite{nonbin3}.
The submatrix $A$ represents shift errors, such that $A_{(i,j)}=a$ if $X(a)$ occurs in
the $j$-th tensor component of the $i$-th error operator in the set of generators.
Similarly, the submatrix $B$ represents phase errors. 
\begin{example}\label{ex:stabilizer2}
The matrix corresponding to the stabilizer state in Example~\ref{ex:stabilizer} is
\[
(A \mid B) = 
\left(
\begin{array}{cccc|cccc}
1 & 1 & 0 & 1 & 0 & 2 & 0 & 0 \\
1 & 2 & 1 & 1 & 1 & 0 & 1 & 0 \\
0 & 2 & 1 & 0 & 0 & 2 & 1 & 2 \\
1 & 2 & 2 & 2 & 0 & 2 & 0 & 2 
\end{array}
\right).
\]
\end{example}
The matrix $(A \mid B)$ generates a code $\mathcal{C}$, and this code is a representation of a
stabilizer state. The fact that a stabilizer is an Abelian group translates into the requirement that
$\mathcal{C}$ must be \emph{self-dual} with respect to a \emph{symplectic inner product}, i.e.,
\[
(a \mid b) * (a' \mid b') = \tr_{m/p}(b \cdot a' - b' \cdot a) = 0, \quad 
\forall (a \mid b), (a' \mid b') \in \mathcal{C}.
\]
We define the \emph{symplectic weight} of a codeword $(a \mid b) \in \mathcal{C}$ as the number of
positions $i$ where $a_i$, $b_i$, or both are nonzero. (This is the same as the weight
of the corresponding Pauli error operator.)

We can also map the linear code of length $2n$ defined above to an additive code
over $\mathbb{F}_{m^2}$ of length $n$. The representation of binary stabilizer codes as self-dual additive codes 
over $\mathbb{F}_4$ was first demonstrated by Calderbank et~al.~\cite{calderbank}, and generalized to
qudits by Ashikhmin and Knill~\cite{nonbin3}, and by Ketkar et~al.~\cite{nonbin2}.
An \emph{additive} code, $\mathcal{C}$, over $\mathbb{F}_{m^2}$ of length~$n$ is defined as an
$\mathbb{F}_m$-linear subgroup of $\mathbb{F}_{m^2}^n$. The code $\mathcal{C}$ contains $m^n$ codewords,
and can be defined by an $n \times n$ generator matrix, $C$, with entries from $\mathbb{F}_{m^2}$,
such that any $\mathbb{F}_m$-linear combination of rows from $C$ is a codeword.\footnote{For additive codes over 
$\mathbb{F}_4$, each codeword is a sum of rows of the generator matrix. However, we also use the 
name ``additive code'' in this more general case.}
To get from the stabilizer representation $(A \mid B)$ to the generator matrix $C$, we simply take
$C = A + \omega B$, where $\omega$ is a primitive element of $\mathbb{F}_{m^2}$.
The code $\mathcal{C}$ will be self-dual, $\mathcal{C} = \mathcal{C}^\perp$, where the dual is defined
with respect to the \emph{Hermitian trace inner product}, $\mathcal{C}^\perp = \{ \boldsymbol{u} \in 
\mathbb{F}_{m^2}^n \mid \boldsymbol{u}*\boldsymbol{c}=0 \text{ for all } \boldsymbol{c} \in \mathcal{C} \}$.
When $m=p$ is prime, the Hermitian trace inner product of two vectors over $\mathbb{F}_{p^2}$ of length~$n$, 
$\boldsymbol{u} = (u_1,u_2,\ldots,u_n)$ and $\boldsymbol{v} = (v_1,v_2,\ldots,v_n)$, is given by
\[
\boldsymbol{u} * \boldsymbol{v} = \tr_{p^2/p}(\boldsymbol{u} \cdot \boldsymbol{v}^p) 
= \boldsymbol{u} \cdot \boldsymbol{v}^p - \boldsymbol{u}^p \cdot \boldsymbol{v} 
= \sum_{i=1}^n (u_i v_i^p - u_i^p v_i),
\]
When $m=p^r$ is not a prime, we use a modification of the Hermitian trace inner product~\cite{nonbin2},
\[
\boldsymbol{u} * \boldsymbol{v} = 
\tr_{m/p}\left( \frac{\boldsymbol{u} \cdot \boldsymbol{v}^m - \boldsymbol{u}^m \cdot \boldsymbol{v} }{\omega - \omega^m} \right),
\]
where $\omega$ is a primitive element of $\mathbb{F}_{m^2}$.

The \emph{Hamming weight} of a codeword $\boldsymbol{u} \in \mathcal{C}$, denoted $\wt(\boldsymbol{u})$,
is the number of nonzero components of $\boldsymbol{u}$.
The \emph{Hamming distance} between $\boldsymbol{u}$ and $\boldsymbol{v}$
is $\wt(\boldsymbol{u} - \boldsymbol{v})$.
The \emph{minimum distance} of the code $\mathcal{C}$ is the minimal Hamming distance
between any two distinct codewords of $\mathcal{C}$. Since $\mathcal{C}$ is an additive code,
the minimum distance is also given by the smallest nonzero weight of any codeword in $\mathcal{C}$.
A code over $\mathbb{F}_{m^2}$ with minimum distance~$d$ is called an $(n,m^n,d)$ code.
The \emph{weight distribution} of the code $\mathcal{C}$ is the sequence
$(A_0, A_1, \ldots, A_n)$, where $A_i$ is the number of codewords of weight~$i$.
The \emph{weight enumerator} of $\mathcal{C}$ is the polynomial
\[
W(x,y) = \sum_{i=0}^n A_i x^{n-i} y^i
\]
For an additive code over $\mathbb{F}_{m^2}$, all $A_i$ must be divisible by $m-1$.

\begin{example}
The stabilizer state in Example~\ref{ex:stabilizer} corresponds to the following generator matrix of
a self-dual additive $(4,3^4,3)$ code.
\[
C = 
\left(
\begin{array}{cccc}
1 & 1+2\omega & 0 & 1  \\
1+\omega & 2 & 1+\omega & 1  \\
0 & 2+2\omega & 1+\omega & 2\omega  \\
1 & 2+2\omega & 2 & 2+2\omega   
\end{array}
\right).
\]
\end{example}

We define two self-dual additive codes, $\mathcal{C}$ and $\mathcal{C}'$ over $\mathbb{F}_{m^2}$, to be
\emph{equivalent} if the codewords of $\mathcal{C}$ can be mapped onto the codewords 
of $\mathcal{C}'$ by a map that preserves the properties of the code, including self-duality.
A permutation of coordinates, or columns of a generator matrix, is such a map.
Other operations can also be applied to the coordinates of $\mathcal{C}$. Let each element 
$a + \omega b \in \mathbb{F}_{m^2}$ be represented as $\binom{a}{b} \in \mathbb{F}_m^2$. 
We can then premultiply this element by a $2 \times 2$ matrix. (We could equivalently have 
applied transformations to pairwise columns of the $2n \times n$ matrix $(A \mid B)$.) 
It was shown by Rains~\cite{nonbin1} that by applying matrices
from the \emph{symplectic group} $\symp_2(m)$ to each coordinate, we preserve the properties of the code.
(This group contains all $2 \times 2$ matrices with elements in $\mathbb{F}_m$ and determinant one.)
For self-dual additive codes over $\mathbb{F}_4$, these symplectic operations can be represented more
simply as multiplication by nonzero elements from $\mathbb{F}_4$ and conjugation of coordinates.
(Conjugation of elements in $\mathbb{F}_{p^2}$ maps $x$ to $x^p$.)
Combined, there are six possible transformations that are equivalent to the six permutations of the 
elements $\{1,\omega,\omega^2\}$ in the coordinate.
The corresponding symplectic group is
\[
\symp_2(2) = \left<A_1=\begin{pmatrix}0&1\\1&1\end{pmatrix}, A_2=\begin{pmatrix}1&1\\0&1\end{pmatrix} \right>,
\]
where $A_1$ represents multiplication by $\omega$ and $A_2$ represents conjugation.
Including coordinate permutations, there are a total of $6^n n!$ maps for a code of length $n$.

For codes over $\mathbb{F}_9$, we observe that $\symp_2(3)$ is a group of order 24 generated by
\[
\symp_2(3) = \left<A_1=\begin{pmatrix}1&1\\1&2\end{pmatrix}, A_2=\begin{pmatrix}1&1\\0&1\end{pmatrix} \right>,
\]
where $A_1$ represents multiplication by $\omega^2$ and $A_2$ represents the map $a + \omega b \mapsto
a + b + \omega b$. By taking powers of $A_1$, we see that we are allowed to multiply a coordinate 
by $x \in \mathbb{F}_9$ only if $x \overline{x} = 1$. However, if we also conjugate the coordinate, we may
multiply by $x \in \mathbb{F}_9$ where $x \overline{x} = 2$. Note that conjugation on its own is not allowed.
The 8 operations just described may be combined with the operations represented by $A_2$ and $A_2^2$ to
give a total of 24 operations. In all there are $24^n n!$ maps that take a self-dual additive code
over $\mathbb{F}_9$ to an equivalent code. In general, for codes over $\mathbb{F}_{m^2}$, the number of maps
is $|\symp_2(m)|^n n!$.

A transformation that maps $\mathcal{C}$ to itself is called an \emph{automorphism} of $\mathcal{C}$.
All automorphisms of $\mathcal{C}$ make up an \emph{automorphism group}, denoted $\Aut(\mathcal{C})$. 
The number of distinct codes equivalent to a self-dual additive code over $\mathbb{F}_{m^2}$, $\mathcal{C}$, is then
given by $\frac{|\symp_2(m)|^n n!}{|\Aut(\mathcal{C})|}$.
The \emph{equivalence class} of $\mathcal{C}$ contains all codes that are equivalent to $\mathcal{C}$.
By adding the sizes of all equivalence classes of codes of length~$n$, we find the total number of distinct codes
of length~$n$, denoted $T_n$. The number $T_n$ is also given by a \emph{mass formula}. The mass formula
for self-dual additive codes over $\mathbb{F}_4$ was found by Höhn~\cite{hohn}. This result is easily generalized
to $\mathbb{F}_{m^2}$.

\begin{theorem}\label{mass}
\[
T_n = \prod_{i=1}^{n} (m^i+1) = \sum_{j=1}^{t_n} \frac{|\symp_2(m)|^n n!}{|\Aut(\mathcal{C}_j)|},
\]
where $t_n$ is the number of equivalence classes of codes of length~$n$, and $\mathcal{C}_j$
is a representative from each equivalence class.
\end{theorem}
\begin{proof}
Let $M(n,k)$ be the total number of self-orthogonal $(n,m^k)$ codes. One such code, $\mathcal{C}$,
can be extended to a self-orthogonal $(n,m^{k+1})$ code in $m^{2(n-k)}-1$ ways by adding
an extra codeword from $C^\perp$. Each $(n,m^{k+1})$ code can be obtained in this way
from $m^{2(k+1)}-1$ different $(n,m^k)$ codes. It follows that
\[
M(n,k+1) = M(n,k) \frac{m^{2(n-k)}-1}{m^{2(k+1)}-1}.
\]
Starting with $M(n,0)=1$, the recursion gives us the number of self-dual $(n,m^n)$ codes,
\[
M(n,n) = \prod_{i=0}^{n-1} \frac{m^{2(n-k)}-1}{m^{2(k+1)}-1} = \prod_{i=1}^{n} (m^i+1).\qedhere
\]
\end{proof}

By assuming that all codes of length $n$ have a trivial automorphism group, we get
the following lower bound on $t_n$, the total number of inequivalent codes.
Note that when $n$ is large,
most codes have a trivial automorphism group, so the tightness of the bound increases with $n$.
Also note that this bound is much tighter than a bound that was derived from results in graph theory 
by Bahramgiri and Beigi~\cite{newlc}.

\begin{theorem}\label{massbound}
\[
t_n \ge \left\lceil\frac{c\prod_{i=1}^n (m^i+1)}{|\symp_2(m)|^n n!}\right\rceil,
\]
where $c=1$ if $m$ is even, and $c=2$ if $m$ is odd.
\end{theorem}
\begin{proof}
When $m$ is even, the trivial automorphism group includes only the identity permutation, and the 
result follows from Theorem~\ref{mass}.
When $m=p^r$ is odd, where $p$ is a prime, the trivial automorphism group also contains 
the transformation that applies the symplectic operation 
$\begin{pmatrix}p-1&0\\0&p-1\end{pmatrix}$ to all coordinates. This operation is equivalent 
to multiplying each codeword by $p-1$, and will therefore map an additive code to itself.
\end{proof}

It follows from the \emph{quantum singleton bound}~\cite{knill,nonbin1} that any self-dual additive code
must satisfy $2d \le n + 2$. A tighter bound for codes over $\mathbb{F}_4$ was given by Calderbank et~al.~\cite{calderbank}.
Codes that satisfy the singleton bound with equality are known as \emph{maximum distance separable (MDS) codes}. 
Self-dual MDS codes must have even length, and MDS codes of length two are trivial and exist for all alphabets. 
The only non-trivial MDS code over $\mathbb{F}_4$ is the $(6,2^6,4)$ \emph{Hexacode}.
Ketkar et~al.~\cite[Thm.~63]{nonbin2} proved that a self-dual additive $(n,m^n,d)$ MDS code must 
satisfy $n \le m^2 + d - 2 \le 2 m^2 -2$. If the famous MDS conjecture holds, then $n \le m^2 + 1$,
or $n \le m^2 + 2$ when $m$ is even and $d=4$ or $d=m^2$.
Grassl, Rötteler, and Beth~\cite{grasslmds} showed that MDS codes of length $n \le m + 1$ always exist.

Self-dual \emph{linear} codes over $\mathbb{F}_{m^2}$ are a subset of the self-dual additive codes.
Only additive codes that satisfy certain constraints can be linear. Such constraints 
for codes over $\mathbb{F}_4$ were described by Van den Nest~\cite{nestthesis} and 
by Glynn et~al.~\cite{glynnbook}. An obvious constraint is that all coefficients of the 
weight enumerator, except $A_0$, of a linear code must be divisible by $m^2-1$, whereas for an additive 
code they need only be divisible by $m-1$.

\section{Correspondence to weighted graphs}\label{sec:graph}

A \emph{graph} is a pair $G=(V,E)$ where $V$ is a set of \emph{vertices} and
$E \subseteq V \times V$ is a set of \emph{edges}.
Let an \emph{$m$-weighted graph} be a triple $G=(V,E,W)$ where $W$ is a set of weights from $\mathbb{F}_m$. 
Each edge has an associated non-zero weight. (An edge with weight zero is the same as a non-edge.)
An $m$-weighted graph with $n$ vertices can be represented by an $n \times n$ \emph{adjacency matrix} $\Gamma$, 
where the element $\Gamma_{(i,j)} = W({\{i,j\}})$ if $\{i,j\} \in E$, and $\Gamma_{(i,j)} = 0$ otherwise.
We will only consider \emph{simple} \emph{undirected} graphs whose adjacency matrices are symmetric with 
all diagonal elements being 0. The \emph{neighbourhood} of $v \in V$, denoted $N_v \subset V$, 
is the set of vertices connected to $v$ by an edge. The number of vertices adjacent to $v$, $|N_v|$,
is called the \emph{degree} of $v$. The \emph{induced subgraph} of $G$ on $U \subseteq V$ 
contains vertices $U$ and all edges from $E$ whose endpoints are both in $U$.
The \emph{complement} of a 2-weighted graph $G$ is found by replacing $E$ with 
$V \times V - E$, i.e., the edges in $E$ are changed to non-edges, and the non-edges to edges.
Two graphs $G=(V,E)$ and $G'=(V,E')$ are \emph{isomorphic} if and only if
there exists a permutation $\pi$ of $V$ such that $\{u,v\} \in E \iff \{\pi(u), \pi(v)\}
\in E'$. We also require that weights are preserved, i.e., $W_{\{u,v\}} = W_{\{\pi(u), \pi(v)\}}$.
A \emph{path} is a sequence of vertices, $(v_1,v_2,\ldots,v_i)$, such that
$\{v_1,v_2\}, \{v_2,v_3\},$ $\ldots, \{v_{i-1},v_{i}\} \in E$.
A graph is \emph{connected} if there is a path from any vertex to any other vertex in the graph.
A \emph{complete graph} is a graph where all pairs of vertices are connected by an edge. A \emph{clique}
is a complete subgraph.

\begin{definition}
A \emph{graph code} is an additive code over $\mathbb{F}_{m^2}$ that has a generator matrix of the form
$C = \Gamma + \omega I$, where $I$ is the identity matrix, $\omega$ is a primitive element of $\mathbb{F}_{m^2}$,
and $\Gamma$ is the adjacency matrix of a simple undirected $m$-weighted graph.
\end{definition}

\begin{theorem}\label{th:graph}
Every self-dual additive code over $\mathbb{F}_{m^2}$ is equivalent to a graph code.
\end{theorem}
\begin{proof}
The generator matrix, $C$, of a self-dual additive code over $\mathbb{F}_{m^2}$
corresponds to an $n \times 2n$ matrix $(A\mid B)$ with elements from $\mathbb{F}_m$, such that $C = A + \omega B$.
We must prove that an equivalent code is generated by $(\Gamma\mid I)$, 
where $I$ is the identity matrix and $\Gamma$ is the adjacency matrix of a simple 
undirected $m$-weighted graph.
A basis change can be accomplished by $(A'\mid B') = M(A\mid B)$, where $M$ is an $n \times n$ 
invertible matrix with elements from $\mathbb{F}_m$.
If $B$ has full rank, the solution is simple, since $B^{-1} (A\mid B) = (\Gamma'\mid I)$.
We obtain $(\Gamma \mid  I)$ after changing the diagonal elements of $\Gamma'$ to 0, by 
appropriate symplectic transformations.
Any two rows of $(\Gamma\mid I)$ will be orthogonal with respect to the symplectic inner product,
which means that $\Gamma I^{\text{T}} - I \Gamma^{\text{T}} = 0$, and it follows that
$\Gamma$ will always be a symmetric matrix.
In the case where $B$ has rank $k<n$, we can perform a basis change to get
\[
(A'\mid B') = 
\left(\begin{array}{c|c}A_1 & B_1 \\ A_2 & \boldsymbol{0} \end{array}\right),
\]
where $B_1$ is a $k \times n$ matrix with full rank, and  $A_1$ also has size $k \times n$.
Since the row-space of $(A'\mid B')$ defines a self-dual code,
and $B'$ contains an all-zero row, it must be true that $A_2 B_1^{\text{T}} = \boldsymbol{0}$.
$A_2$ must have full rank, and the row space
of $B_1$ must be the orthogonal complement of the row space of $A_2$.
We assume that $B_1 = (B_{11} \mid  B_{12})$ where $B_{11}$ is a $k \times k$ invertible
matrix. We also write $A_2 = (A_{21} \mid  A_{22})$ where $A_{22}$ has size
$(n-k) \times (n-k)$. 
Assume that there exists an $\boldsymbol{x} \in \mathbb{F}_m^{n-k}$ such that $A_{22} \boldsymbol{x}^{\text{T}} = 0$.
Then the vector $\boldsymbol{v} = (0, \ldots, 0, \boldsymbol{x})$ of length~$n$
satisfies $A_2 \boldsymbol{v}^{\text{T}} = 0$.
Since the row space of $B_1$ is the orthogonal complement of the row space of $A_2$,
we can write $\boldsymbol{v} = \boldsymbol{y} B_1$ for some $\boldsymbol{y} \in \mathbb{F}_m^k$.
We see that $\boldsymbol{y} B_{11}  = 0$, and since $B_{11}$ has full rank, it must therefore be true that
$\boldsymbol{y} = 0$. This means that $\boldsymbol{x} = 0$, which proves that
$A_{22}$ is an invertible matrix.
Two of the symplectic operations that we can apply to columns of a generator matrix
are $\begin{pmatrix}0&m-1\\1&0\end{pmatrix}$ and $\begin{pmatrix}0&1\\m-1&0\end{pmatrix}$.
This means that we can interchange column~$i$ of $A'$ and column~$i$ of $B'$ if we
also multiply one of the columns by $m-1$.
In this way we swap the $i$-th columns of $A'$ and $B'$ for $k < i \le n$ to get
$(A''\mid B'')$. Since $B_{11}$ and $A_{22}$ are invertible, $B''$ must also be an invertible matrix.
We then find $B''^{-1} (A''\mid B'') = (\Gamma\mid I)$, and set all diagonal elements of $\Gamma$ to 0
by symplectic transformations.
\end{proof}

\begin{figure}
 \centering
 \includegraphics[height=.20\linewidth]{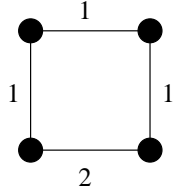}
 \caption{Graph Representation of the $(4,3^4,3)$ Code}\label{4d3graph}
\end{figure}

\begin{example}\label{ex:stabilizer3}
The matrix from Example~\ref{ex:stabilizer2} can be transformed into the following
matrix, using the method given in the proof of Theorem~\ref{th:graph}.
\[
(\Gamma \mid I) = 
\left(
\begin{array}{cccc|cccc}
0 & 1 & 1 & 0 & 1 & 0 & 0 & 0 \\
1 & 0 & 0 & 1 & 0 & 1 & 0 & 0 \\
1 & 0 & 0 & 2 & 0 & 0 & 1 & 0 \\
0 & 1 & 2 & 0 & 0 & 0 & 0 & 1 
\end{array}
\right).
\]
This means that the stabilizer state from Example~\ref{ex:stabilizer} is equivalent to the graph code
generated by $C = \Gamma + \omega I$. The graph defined by $\Gamma$ is depicted in Fig.~\ref{4d3graph}.
\end{example}

Note that Theorem~\ref{th:graph} is a generalization of the same theorem for codes over 
$\mathbb{F}_4$~\cite{selfdualgf4}, which was proved by Van~den~Nest~et~al.~\cite{nest}.
The fact that stabilizer codes can be represented by graphs was also shown by Schlingemann 
and Werner~\cite{schlingemann} and by Grassl, Klappenecker, and Rötteler~\cite{grasslpaper}.

We have seen that every $m$-weighted graph represents a self-dual additive code over $\mathbb{F}_{m^2}$,
and that every self-dual additive code over $\mathbb{F}_{m^2}$ can be represented by an $m$-weighted graph.
It follows that we can, without loss of generality,
restrict our study to codes with generator matrices of the form $\Gamma + \omega I$,
where $\Gamma$ is an adjacency matrix of an unlabeled simple undirected $m$-weighed graph.

\section{Graph equivalence and code equivalence}

Swapping vertex~$i$ and vertex $j$ of a graph with adjacency matrix $\Gamma$ can be accomplished by
exchanging column~$i$ and column~$j$ of $\Gamma$ and then exchanging row~$i$ and row~$j$ of $\Gamma$.
We call the resulting matrix $\Gamma'$. Exactly the same column and row operations map
$\Gamma + \omega I$ to $\Gamma' + \omega I$, which are generator matrices for equivalent codes.
It follows that two codes are equivalent if their corresponding graphs are isomorphic.
However, the symplectic transformations that map a code to an equivalent code do not in general produce
isomorphic graphs, but we will see that they can be described as graph operations.

It is known that two self-dual additive codes over $\mathbb{F}_4$ are equivalent
if and only if their corresponding graphs are equivalent, up to isomorphism, with respect to a sequence
of \emph{local complementations}~\cite{bouchet4,glynnbook,nestthesis,nest}.
We have previously used this fact to devise a graph-based algorithm with which we classified all self-dual
additive codes over $\mathbb{F}_4$ of length up to 12~\cite{selfdualgf4}.

\begin{definition}[\cite{bouchet4}]
Given a graph $G=(V,E)$ and a vertex $v \in V$, let $N_v \subset V$ be the neighbourhood of $v$.
\emph{Local complementation} (LC) on $v$ transforms $G$ into $G * v$ by replacing the induced subgraph 
of $G$ on $N_v$ by its complement.
\end{definition}

\begin{theorem}[\cite{bouchet4,glynnbook,nestthesis,nest}]\label{th:lc}
Two self-dual additive codes over $\mathbb{F}_4$, $\mathcal{C}$ and $\mathcal{C}'$,
with graph representations $G$ and $G'$, are equivalent if and only if
there is a finite sequence of not necessarily distinct vertices
$(v_1, v_2, \ldots, v_i)$, such that $G * v_1 * v_2 * \cdots * v_i$ is
isomorphic to $G'$.
\end{theorem}

\begin{figure}
 \centering
 \subfloat[The Graph $G$]
 {\hspace{5pt}\includegraphics[height=.20\linewidth]{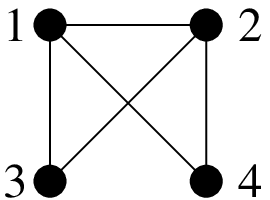}\hspace{5pt}\label{fig:lcexample1}}
 \quad
 \subfloat[The Graph $G*1$]
 {\hspace{5pt}\includegraphics[height=.20\linewidth]{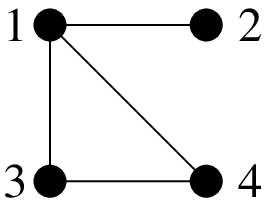}\hspace{5pt}\label{fig:lcexample2}}
 \caption{Example of Local Complementation}\label{fig:lcexample}
\end{figure}

The LC operation can be generalized to weighted graphs, and it was first shown
by Bahramgiri and Beigi~\cite{newlc} that the equivalence of nonbinary stabilizer 
states over $\mathbb{F}_m$, i.e., self-dual additive codes over $\mathbb{F}_{m^2}$, can
be described in terms of graph operations.\footnote{Bahramgiri and Beigi~\cite{newlc} only
state their theorem for $\mathbb{F}_m$ where $m$ is prime, but the result holds for any finite field,
as their proof does not depend on $m$ being prime.}

\begin{definition}[\cite{newlc}]\label{weightswap}
Given an $m$-weighted graph $G=(V,E,W)$ and a vertex $v \in V$,
\emph{weight shifting} on $v$ by $a \in \mathbb{F}_m \setminus \{0\}$ transforms $G$ into $G \circ_a v$
by multiplying the weight of each edge incident on $v$ by $a$.
\end{definition}

\begin{figure}
 \centering
 \subfloat[The Graph $G$]
 {\hspace{5pt}\includegraphics[height=.20\linewidth]{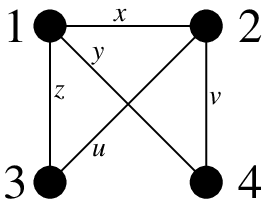}\hspace{5pt}\label{fig:wsexample1}}
 \quad
 \subfloat[The Graph $G \circ_a 1$]
 {\hspace{5pt}\includegraphics[height=.20\linewidth]{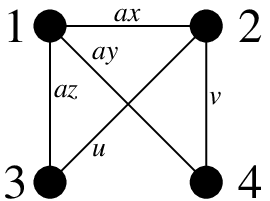}\hspace{5pt}\label{fig:wsexample2}}
 \caption{Example of Weight Shifting}\label{fig:wsexample}
\end{figure}

\begin{definition}[\cite{newlc}]\label{generallc}
Given an $m$-weighted graph $G=(V,E,W)$ and a vertex $v \in V$,
\emph{generalized local complementation} on $v$ by $a \in \mathbb{F}_m \setminus \{0\}$ transforms $G$ 
into $G *_a v$. Let $\Gamma$ and $\Gamma'$ be the adjacency matrices of $G$ and $G *_a v$, respectively.
Then $\Gamma'_{(i,j)} = \Gamma_{(i,j)} + a\Gamma_{(v,i)}\Gamma_{(v,j)}$, for all $i \ne j$, and 
$\Gamma'_{(i,i)} = 0$ for all $i$.
\end{definition}

\begin{figure}
 \centering
 \subfloat[The Graph $G$]
 {\hspace{5pt}\includegraphics[height=.20\linewidth]{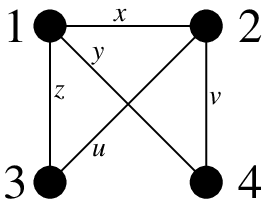}\hspace{5pt}\label{fig:glcexample1}}
 \quad
 \subfloat[The Graph $G *_a 1$]
 {\hspace{5pt}\includegraphics[height=.20\linewidth]{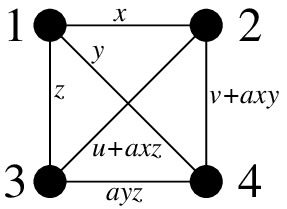}\hspace{5pt}\label{fig:glcexample2}}
 \caption{Example of Generalized Local Complementation}\label{fig:glcexample}
\end{figure}

\begin{theorem}[\cite{newlc}]\label{th:genlc}
Two self-dual additive codes over $\mathbb{F}_{m^2}$, $\mathcal{C}$ and $\mathcal{C}'$,
with graph representations $G$ and $G'$, are equivalent if and only if
we get a graph isomorphic to $G'$ by applying some finite sequence of 
weight shifts and generalized local complementations to $G$.
\end{theorem}

A proof of Theorem~\ref{th:genlc} was given by by Bahramgiri and Beigi~\cite{newlc}, as
a generalization of the proof given by Van den Nest et al.~\cite{nest} for
self-dual additive codes over $\mathbb{F}_4$.

\begin{definition}
The \emph{LC orbit} of a weighted graph $G$ is the set of all non-isomorphic graphs that
can be obtained by performing any sequence of weight shifts and generalized LC operations on~$G$.
\end{definition}

\begin{theorem}
The minimum distance of a self-dual additive $(n,m^n,d)$ code is equal to $\delta + 1$, where
$\delta$ is the minimum vertex degree over all graphs in the associated LC orbit.
\end{theorem}
\begin{proof}
A vertex with degree $d-1$ in the LC orbit corresponds to a codeword of weight~$d$,
and we will now show that such a vertex always exists. Choose any graph representation of the code
and let $G = (\Gamma \mid I)$ be the corresponding generator matrix.
Find a codeword $c$ of weight~$d$ generated by $G$. 
Let the $i$-th row of $G$ be one of the rows that $c$ is linearly dependent on.
Apply symplectic transformations to the coordinates of the code such that $c$ is mapped to 
$c'$ with $1$ in coordinate $n+i$, and with $0$ in all other of the last $n$ coordinates.
Since we do not care about changes in the corresponding first $n$ coordinates, 
as long as the symplectic weight of $c$ is preserved, there will always be transformations that achieve this.
Apply the same transformations to the columns of $G$, and then replace the $i$-th row with $c'$, to get~$G'$. 
Note that the right half of $G'$ still has full rank, so we can transform $G'$ into a matrix
of the form $(\Gamma' \mid I)$ by Gaussian elimination, where the symplectic weight of the $i$-th row is $d$.
Finally, we set all diagonal elements of $\Gamma'$ to zero by appropriate symplectic transformations. 
Vertex $i$ of the graph with adjacency matrix $\Gamma'$ has degree $d-1$.
\end{proof}

\section{Classification}\label{sec:class}

It follows from Theorem~\ref{th:genlc} that two self-dual additive codes over $\mathbb{F}_{m^2}$
are equivalent if and only if their graph representations are in the same LC orbit.
The LC orbit of a graph can easily be generated by a recursive algorithm.
We have used the program \emph{nauty}~\cite{nauty} to check for graph isomorphism.

Let $\boldsymbol{G}_{n,m}$ be the set consisting of all non-isomorphic simple undirected connected 
$m$-weighted graphs on $n$ vertices.
Note that connected graphs correspond to \emph{indecomposable} codes.
A code is decomposable if it can be written as the \emph{direct sum} of two smaller codes.
For example, let $\mathcal{C}$ be an $(n,m^n,d)$ code and $\mathcal{C}'$ an $(n',m^{n'},d')$ code. The
direct sum, $\mathcal{C} \oplus \mathcal{C}' = \{u||v \mid u \in \mathcal{C}, v \in \mathcal{C}'\}$,
where $||$ means concatenation, is an $({n+n'},m^{n+n'},\min\{d,d'\})$ code.
It follows that all decomposable codes of length~$n$ can be classified easily once
all indecomposable codes of length less than $n$ are known.

The set of all distinct LC orbits of connected $m$-weighted graphs on $n$ vertices is a partitioning of
$\boldsymbol{G}_{n,m}$ into $i_{n,m}$ disjoint sets.
$i_{n,m}$ is also the number of indecomposable self-dual additive codes over $\mathbb{F}_{m^2}$ of length~$n$,
up to equivalence. Let $\boldsymbol{L}_{n,m}$ be a set containing one representative from each
LC orbit of connected $m$-weighted graphs on $n$ vertices.
The simplest algorithm for finding such sets of representatives is to start with the set 
$\boldsymbol{G}_{n,m}$ and generate LC orbits of its members until we have a partitioning of $\boldsymbol{G}_{n,m}$.
The following more efficient technique is based on a method described by Glynn et~al.~\cite{glynnbook}.
Let the $m^n-1$ \emph{extensions} of an $m$-weighted graph on $n$ vertices be formed by
adding a new vertex and joining it to all possible combinations of at least one of the old vertices, using
all possible combinations of edge weights.
The set $\boldsymbol{E}_{n,m}$, containing $i_{n-1,m} (m^{n-1}-1)$ graphs,
is formed by making all possible extensions of all graphs in $\boldsymbol{L}_{n-1,m}$.

\begin{theorem}\label{thm:extend}
$\boldsymbol{L}_{n,m} \subset \boldsymbol{E}_{n,m}$, i.e.,
the set $\boldsymbol{E}_{n,m}$ will contain at least one representative from each
LC orbit of connected $m$-weighted graphs on $n$ vertices.
\end{theorem}
\begin{proof}
Let $G=(V,E,W) \in \boldsymbol{G}_{n,m}$, and choose any subset $U \subset V$ of $n-1$ vertices.
By doing weight shifts and generalized LC operations on vertices in $U$, we can transform the induced subgraph of 
$G$ on $U$ into one of the graphs in $\boldsymbol{L}_{n-1,m}$ that were extended when $\boldsymbol{E}_{n,m}$ 
was constructed. It follows that for all $G \in \boldsymbol{G}_{n,m}$, some graph in the LC orbit of $G$ must be 
part of $\boldsymbol{E}_{n,m}$.
\end{proof}

The set $\boldsymbol{E}_{n,m}$ will be much smaller than $\boldsymbol{G}_{n,m}$, so it will be more
efficient to search for a set of LC orbit representatives within $\boldsymbol{E}_{n,m}$. 
Another fact that simplifies our classification algorithm is that weight shifting 
and generalized local complementation commute. This means that to generate the LC orbit of a
weighted graph, we may first generate the orbit with respect to generalized local complementation only, 
and then apply weight shifting to the resulting set of graphs.

Using the described techniques, we were able to classify all self-dual additive codes over 
$\mathbb{F}_9$, $\mathbb{F}_{16}$, and $\mathbb{F}_{25}$ up to lengths 8, 6, and 6, respectively.
Table~\ref{tab:orbitsindecomp} gives the values of $i_{n,m}$, the number of distinct LC orbits
of connected $m$-weighted graphs on $n$ vertices, which is also the number of inequivalent indecomposable 
self-dual additive codes over $\mathbb{F}_{m^2}$ of length~$n$. 
The total number of inequivalent codes of length~$n$, $t_n$,
is shown in Table~\ref{tab:orbitsall} together with lower bounds derived from Theorem~\ref{massbound}.
The numbers $t_n$ are easily derived from the numbers $i_n$ by using the \emph{Euler transform}~\cite{sloane2},
\begin{eqnarray*}
c_n &=& \sum_{d|n} d i_d\\
t_1 &=& c_1\\
t_n &=& \frac{1}{n}\left( c_n + \sum_{k=1}^{n-1} c_k t_{n-k} \right).
\end{eqnarray*}
Tables~\ref{tab:distance3}, \ref{tab:distance4}, and \ref{tab:distance5} list by minimum distance the 
numbers of indecomposable codes over $\mathbb{F}_9$, $\mathbb{F}_{16}$, and $\mathbb{F}_{25}$.
A database containing one representative from each equivalence class is available 
at \url{http://www.ii.uib.no/~larsed/nonbinary/}.
For the classification of self-dual additive codes over $\mathbb{F}_4$,
we refer to previous work~\cite{selfdualgf4}, and the web page \url{http://www.ii.uib.no/~larsed/vncorbits/}.

\begin{table}
\centering
\caption{Number ($i_{n,m}$) of Indecomposable Codes of Length~$n$ over $\mathbb{F}_{m^2}$}
\label{tab:orbitsindecomp}
\begin{tabular}{ccccccc}
\toprule
$n$ & $i_{n,2}$ & $i_{n,3}$ & $i_{n,4}$ & $i_{n,5}$ \\
\midrule
 1 & 1 &   1 &  1 &  1 \\
 2 & 1 &   1 &  1 &  1 \\ 
 3 & 1 &   1 &  1 &  1 \\
 4 & 2 &   3 &  3 &  3 \\
 5 & 4 &   5 &  6 &  7 \\
 6 & 11 & 21 & 25 & 38 \\
 7 & 26 & 73 & & \\
 8 & 101 & 659 &  & \\
 9 & 440 &   &  & \\
10 & 3,132 & & \\
11 & 40,457 & & & \\
12 & 1,274,068 & & & \\
\bottomrule
\end{tabular}
\end{table}

\begin{table}
\centering
\caption{Total Number ($t_{n,m}$) of Codes of Length~$n$ over $\mathbb{F}_{m^2}$}
\label{tab:orbitsall}
\begin{tabular}{ccccccc}
\toprule
$n$ & $t_{n,2}$ & $t_{n,3}$ & $t_{n,4}$ & $t_{n,5}$ \\
\midrule
 1 & 1 &     1 &  1 &  1 \\
 2 & 2 &     2 &  2 &  2 \\
 3 & 3 &     3 &  3 &  3 \\
 4 & 6 &     7 &  7 &  7 \\
 5 & 11 &   13 & 14 & 15 \\
 6 & 26 &   39 & 44 & 58 \\
 7 & 59 &  121 &  ? &  ? \\
 8 & 182 & 817 & \emph{$\ge$ 946} & \emph{$\ge$ 21,161} \\
 9 & 675 & \emph{$\ge$ 9,646} & \emph{$\ge$ 458,993} & \emph{$\ge$ 38,267,406} \\
10 & 3,990 & \emph{$\ge$ 2,373,100} & & \\
11 & 45,144 & & & \\
12 & 1,323,363 & & & \\
13 & \emph{$\ge$ 72,573,549} & & & \\
\bottomrule
\end{tabular}
\end{table}

\begin{table}
\centering
\caption{Number of Indecomposable Codes of Length~$n$ and Distance~$d$ over $\mathbb{F}_9$}
\label{tab:distance3}
\begin{tabular}{crrrrrrrrr}
\toprule
$d \backslash n$ & 
\multicolumn{1}{c}{2} & \multicolumn{1}{c}{3} & \multicolumn{1}{c}{4} & \multicolumn{1}{c}{5} &
\multicolumn{1}{c}{6} & \multicolumn{1}{c}{7} & \multicolumn{1}{c}{8} & \multicolumn{1}{c}{9} &
\multicolumn{1}{c}{10}\\
\midrule
2     & 1 & 1 & 2 &  4 & 15 & 51 & 388 &   ? &     ? \\
3     &   &   & 1 &  1 &  5 & 20 & 194 &   ? &     ? \\
4     &   &   &   &    &  1 &  2 &  77 &   ? &     ? \\
5     &   &   &   &    &    &    &     &   4 &     ? \\
6     &   &   &   &    &    &    &     &     &     1 \\
\midrule
All   & 1 & 1 & 3 &  5 & 21 & 73 & 659 &   ? &     ? \\
\bottomrule
\end{tabular}
\end{table}

\begin{table}
\centering
\caption{Number of Indecomposable Codes of Length~$n$ and Distance~$d$ over $\mathbb{F}_{16}$}
\label{tab:distance4}
\begin{tabular}{crrrrrrr}
\toprule
$d \backslash n$ & 
\multicolumn{1}{c}{2} & \multicolumn{1}{c}{3} & \multicolumn{1}{c}{4} & \multicolumn{1}{c}{5} &
\multicolumn{1}{c}{6}\\
\midrule
2     & 1 & 1 & 2 &  4 & 16 \\
3     &   &   & 1 &  2 &  6 \\
4     &   &   &   &    &  3 \\
5     &   &   &   &    &    \\
6     &   &   &   &    &    \\
\midrule
All   & 1 & 1 & 3 &  6 & 25 \\
\bottomrule
\end{tabular}
\end{table}

\begin{table}
\centering
\caption{Number of Indecomposable Codes of Length~$n$ and Distance~$d$ over $\mathbb{F}_{25}$}
\label{tab:distance5}
\begin{tabular}{crrrrrrr}
\toprule
$d \backslash n$ & 
\multicolumn{1}{c}{2} & \multicolumn{1}{c}{3} & \multicolumn{1}{c}{4} & \multicolumn{1}{c}{5} &
\multicolumn{1}{c}{6}\\
\midrule
2     & 1 & 1 & 2 &  4 & 21 \\
3     &   &   & 1 &  3 & 11 \\
4     &   &   &   &    &  6 \\
5     &   &   &   &    &    \\
6     &   &   &   &    &    \\
\midrule
All   & 1 & 1 & 3 &  7 & 38 \\
\bottomrule
\end{tabular}
\end{table}

Note that applying the graph extension technique described previously is equivalent to
\emph{lengthening}~\cite{gaborit} a self-dual additive code. Given an $(n,m^n,d)$ code,
we add a row and column to its generator matrix to obtain an $(n+1,m^{n+1},d')$ code,
where $d' \le d+1$. If follows that given a classification of all codes of length $n$ and minimum distance $d$,
we can classify all codes of length $n+1$ and minimum distance $d+1$. All length 8 codes over
$\mathbb{F}_9$ have been classified as described above. By extending the 77 $(8,3^8,4)$ codes,
we found 4 $(9,3^9,5)$ codes, and from those we obtained a single $(10,3^{10},6)$ code.
Assuming that the MDS conjecture holds, there are no self-dual additive MDS codes over $\mathbb{F}_9$
with length above 10. This would mean that the three MDS codes with parameters $(4,3^4,3)$, 
$(6,3^6,4)$, and $(10,3^{10},6)$ are the only non-trivial self-dual additive MDS codes 
over $\mathbb{F}_9$. The $(6,3^6,4)$ and $(10,3^{10},6)$ are constructed as circulant codes in
Section~\ref{sec:circulant}. A generator matrix for the $(4,3^4,3)$ code is given in Example~\ref{ex:stabilizer3}.
In fact, a $(4,m^4,3)$ code, for any $m \ge 3$, is generated by
\[
\left(
\begin{array}{cccc}
\omega & 1 & 1 & 0 \\
1 & \omega & 0 & 1 \\
1 & 0 & \omega & \alpha \\
0 & 1 & \alpha & \omega   
\end{array}
\right),
\]
where $\alpha \in \mathbb{F}_m \setminus \{0,1\}$.
This code has weight enumerator $W(1,y) = 1 + 4(m^2-1)y^3 + (m^2-3)(m^2-1)y^4$.

There are four $(9,3^9,5)$ codes, all with weight enumerator 
$W(1,y) = 1 + 252 y^5 + 1176 y^6 + 3672 y^7 + 7794 y^8 + 6788 y^9$.
None of these are equivalent to circulant graph codes. The generator matrices are:
\[
\left(
\begin{array}{ccccccccc}
\omega & 2 & 2 & 1 & 0 & 0 & 0 & 2 & 2 \\ 
2 & \omega & 1 & 0 & 0 & 0 & 1 & 2 & 1 \\ 
2 & 1 & \omega & 0 & 0 & 2 & 2 & 1 & 1 \\ 
1 & 0 & 0 & \omega & 2 & 1 & 0 & 0 & 1 \\ 
0 & 0 & 0 & 2 & \omega & 2 & 1 & 1 & 1 \\ 
0 & 0 & 2 & 1 & 2 & \omega & 1 & 1 & 0 \\ 
0 & 1 & 2 & 0 & 1 & 1 & \omega & 0 & 2 \\ 
2 & 2 & 1 & 0 & 1 & 1 & 0 & \omega & 2 \\ 
2 & 1 & 1 & 1 & 1 & 0 & 2 & 2 & \omega 
\end{array}
\right),
\left(
\begin{array}{ccccccccc}
\omega & 2 & 2 & 2 & 2 & 2 & 2 & 2 & 1 \\ 
2 & \omega & 0 & 0 & 0 & 0 & 2 & 1 & 2 \\ 
2 & 0 & \omega & 0 & 2 & 0 & 1 & 0 & 2 \\ 
2 & 0 & 0 & \omega & 1 & 0 & 1 & 2 & 0 \\ 
2 & 0 & 2 & 1 & \omega & 1 & 2 & 2 & 1 \\ 
2 & 0 & 0 & 0 & 1 & \omega & 0 & 1 & 1 \\ 
2 & 2 & 1 & 1 & 2 & 0 & \omega & 2 & 1 \\ 
2 & 1 & 0 & 2 & 2 & 1 & 2 & \omega & 1 \\ 
1 & 2 & 2 & 0 & 1 & 1 & 1 & 1 & \omega 
\end{array}
\right),
\]
\[
\left(
\begin{array}{ccccccccc}
\omega & 2 & 0 & 0 & 0 & 2 & 1 & 2 & 1 \\ 
2 & \omega & 0 & 2 & 1 & 2 & 0 & 1 & 0 \\ 
0 & 0 & \omega & 0 & 1 & 2 & 0 & 1 & 1 \\ 
0 & 2 & 0 & \omega & 0 & 1 & 0 & 1 & 1 \\ 
0 & 1 & 1 & 0 & \omega & 1 & 2 & 2 & 1 \\ 
2 & 2 & 2 & 1 & 1 & \omega & 1 & 2 & 1 \\ 
1 & 0 & 0 & 0 & 2 & 1 & \omega & 1 & 1 \\ 
2 & 1 & 1 & 1 & 2 & 2 & 1 & \omega & 1 \\ 
1 & 0 & 1 & 1 & 1 & 1 & 1 & 1 & \omega
\end{array}
\right),
\left(
\begin{array}{ccccccccc}
\omega & 0 & 2 & 0 & 2 & 2 & 0 & 2 & 2 \\ 
0 & \omega & 2 & 2 & 0 & 0 & 2 & 0 & 1 \\ 
2 & 2 & \omega & 0 & 0 & 1 & 0 & 2 & 0 \\ 
0 & 2 & 0 & \omega & 0 & 0 & 1 & 2 & 1 \\ 
2 & 0 & 0 & 0 & \omega & 0 & 2 & 1 & 1 \\ 
2 & 0 & 1 & 0 & 0 & \omega & 1 & 1 & 2 \\ 
0 & 2 & 0 & 1 & 2 & 1 & \omega & 1 & 0 \\ 
2 & 0 & 2 & 2 & 1 & 1 & 1 & \omega & 1 \\ 
2 & 1 & 0 & 1 & 1 & 2 & 0 & 1 & \omega
\end{array}
\right).
\]
The two $(7,3^7,4)$ codes, the three $(6,4^6,4)$, and five of the six $(6,5^6,4)$ codes
are equivalent to circulant graph codes generated in Section~\ref{sec:circulant}. The last $(6,5^6,4)$
code has weight enumerator $W(1,y)= 1 + 360 y^4 + 3024 y^5 + 12240 y^6$ and generator matrix
\[
\left(
\begin{array}{cccccc}
\omega & 2 & 2 & 1 & 0 & 0 \\ 
2 & \omega & 0 & 1 & 4 & 1 \\ 
2 & 0 & \omega & 0 & 1 & 1 \\ 
1 & 1 & 0 & \omega & 3 & 4 \\ 
0 & 4 & 1 & 3 & \omega & 3 \\ 
0 & 1 & 1 & 4 & 3 & \omega 
\end{array}
\right).
\]

\section{Circulant graph codes}\label{sec:circulant}

It is clearly infeasible to study all self-dual additive codes of lengths much higher than those
classified in the previous section. We therefore restrict our search space to the
$m^{\left\lceil\frac{n-1}{2}\right\rceil}$ codes over $\mathbb{F}_{m^2}$ of length $n$ corresponding to graphs
with \emph{circulant} adjacency matrices. A matrix is circulant if the $i$-th row is equal to the
first row, cyclically shifted $i-1$ times to the right. 
We have performed an exhaustive search of such 
graphs, the result of which is summarized in Table~\ref{tab:circulant}. 
This table shows the highest found minimum distance of self-dual additive codes
over various alphabets. A code with the given minimum distance has been found in our search,
except for the cases marked~$*$, where a better code is obtained in some other way and does not have a
circulant graph representation,\footnote{See the web page \url{http://www.codetables.de/} for details
on how codes over $\mathbb{F}_{4}$ of length 18 and 21 can be obtained.}
and cases marked~$s$, which are not circulant, but obtained by
a trivial \emph{shortening}~\cite{gaborit} of a longer circulant code.
Minimum distances printed in bold font are optimal according to the quantum singleton bound. 
If $n$ is even and the quantum singleton bound is satisfied with equality, we have an MDS code.

\begin{table}
\centering
\caption{Highest Found Minimum Distance of Codes over $\mathbb{F}_{m^2}$ of Length $n$}
\label{tab:circulant}
\begin{tabular}{ccccc}
\toprule
$n\backslash m$ & $2$ & $3$ & $4$ & $5$ \\
\midrule
 2 & $\boldsymbol{2}$ & $\boldsymbol{2}$ & $\boldsymbol{2}$ & $\boldsymbol{2}$ \\
 3 & $\boldsymbol{2}$ & $\boldsymbol{2}$ & $\boldsymbol{2}$ & $\boldsymbol{2}$ \\
 4 & 2 & \skp$\boldsymbol{3}^*$ & \skp$\boldsymbol{3}^*$ &  \skp$\boldsymbol{3}^*$ \\
 5 & $\boldsymbol{3}$ & $\boldsymbol{3}$ & $\boldsymbol{3}$ & $\boldsymbol{3}$ \\
 6 & $\boldsymbol{4}$ & $\boldsymbol{4}$ & $\boldsymbol{4}$ & $\boldsymbol{4}$ \\
 7 & 3 & $\boldsymbol{4}$ & $\boldsymbol{4}$ & $\boldsymbol{4}$ \\
 8 & 4 & 4 & 4 & 4 \\
 9 & 4 & \skp$\boldsymbol{5}^s$ & $\boldsymbol{5}$ & $\boldsymbol{5}$ \\
10 & 4 & $\boldsymbol{6}$ & $\boldsymbol{6}$ & $\boldsymbol{6}$ \\
11 & \skp$5^s$ & 5 & $\boldsymbol{6}$ & $\boldsymbol{6}$ \\
12 & 6 & 6 & 6 & 6 \\
13 & 5 & 6 & 6 & $\boldsymbol{7}$ \\
14 & 6 & 6 & 7 & $\boldsymbol{8}$ \\
15 & 6 & 6 & 7 & 7 \\
16 & 6 & 6 & 8 & 8 \\
17 & 7 & 7 & 8 & $\boldsymbol{9}$ \\
18 & \skp$8^*$ & 8 & 8  & $\boldsymbol{10}$ \\
19 & 7 & 8 &   &  \\
20 & 8 & 8 &   &  \\
21 & \skp$8^*$ & 8 &  &  \\
22 & 8 & 9 &   &  \\
23 & 8 & 9 &   &  \\
24 & 8 & 9 &  &  \\
25 & 8 &  &  &  \\
26 & 8 &  &  &  \\
27 & \skp$9^s$ &  &  &  \\
28 & 10 &  &  & \\
29 & 11 &  &  &  \\
30 & 12 &  &  & \\
\bottomrule
\end{tabular}
\end{table}

We here give the first row of a circulant generator matrix for those codes classified in 
Section~\ref{sec:class} that are equivalent to circulant graph codes.
There is a unique $(6,3^6,4)$ code with weight enumerator $W(1,y) = 1 + 120 y^4 + 240 y^5 + 368 y^6$
generated by $(\omega 01110)$. There are two inequivalent $(7,3^7,4)$ codes generated 
by $(\omega 110011)$ and $(\omega 022220)$, both with weight enumerator
$W(1,y) = 1 + 70 y^4 + 336 y^5 + 812 y^6 + 968 y^7$. There is a unique $(10,3^{10},6)$ code 
with weight enumerator $W(1,y) = 1 + 1680 y^6 + 2880 y^7 + 14040 y^8 + 22160 y^9 + 18288 y^{10}$
generated by $(\omega 012111210)$.
There are three inequivalent $(6,4^6,4)$ codes with weight enumerator $W(1,y)= 1 + 225 y^4 + 1080 y^5 + 2790 y^6$
generated by $(\omega 01110)$, $(\omega 01\alpha 10)$, and $(\omega 01\alpha^2 10)$,
where $\alpha = \omega^5$ is a primitive element of $\mathbb{F}_4$.
There are five inequivalent $(6,5^6,4)$ codes generated by $(\omega 01110)$, $(\omega 01210)$, 
$(\omega 02220)$, $(\omega 10201)$, and $(\omega 12221)$, all with weight enumerator 
$W(1,y)= 1 + 360 y^4 + 3024 y^5 + 12240 y^6$.

For circulant graph codes of higher length that are optimal according to the quantum singleton bound,
we find that all codes of the same length have the same weight enumerator. In the list below, we give
the first row of one generator matrix for each weight enumerator. 
\begin{itemize}
\item $(7,4^7,4)$, 
$(\omega 11\alpha \alpha 11)$,
\[
W(1,y) = 1 + 105 y^4 + 1008 y^5 + 4830 y^6 + 10440 y^7.
\]
\item $(9,4^9,5)$,
$(\omega 001\alpha \alpha 100)$,
\[
W(1,y) = 1 + 378 y^5 + 3780 y^6 + 23220 y^7 + 88155 y^8 + 146610 y^9.
\]
\item $(10,4^{10},6)$,
$(\omega 010\alpha 1\alpha 010)$,
\[
W(1,y) = 1 + 3150 y^6 + 18000 y^7 + 111375 y^8 + 366000 y^9 + 550050 y^{10}.
\]
\item $(11,4^{11},6)$,
$(\omega 00\alpha 1111\alpha 00)$,
\begin{align*}
W(1,y) &= 1 + 1386 y^{6} + 13860 y^{7} + 99495 y^{8} + 505560 y^{9} + 1511598 y^{10} \\
& \quad + 2062404 y^{11}.
\end{align*}
\item $(7,5^7,4)$,
$(\omega 011110)$,
\[
W(1,y) = 1 + 140 y^{4} + 2184 y^{5} + 17080 y^{6} + 58720 y^{7}.
\]
\item $(9,5^9,5)$,
$(\omega 00211200)$,
\[
W(1,y) = 1 + 504 y^5 + 8400 y^6 + 84240 y^7 + 507420 y^8 + 1352560 y^9.
\]
\item $(10,5^{10},6)$,
$(\omega 001222100)$,
\[
W(1,y) = 1 + 5040 y^6 + 54720 y^7 + 508680 y^8 + 2704560 y^9 + 6492624 y^{10}.
\]
\item $(11,5^{11},6)$,
$(\omega 0012222100)$,
\begin{align*}
W(1,y) &= 1 + 1848 y^6 + 31680 y^7 + 370260 y^8 + 2977480 y^9 \\
& \quad + 14282664 y^{10} + 31164192 y^{11}.
\end{align*}
\item $(13,5^{13},7)$,
$(\omega 010011110010)$,
\begin{align*}
W(1,y) &= 1 + 6864 y^7 + 118404 y^8 + 1538680 y^9 + 14867424 y^{10} \\
& \quad + 97222320 y^{11} + 388930776 y^{12} + 718018656 y^{13}.
\end{align*}
\item $(14,5^{14},8)$,
$(\omega 1011331331101)$,
\begin{align*}
W(1,y) &= 1 + 72072 y^8 + 816816 y^9 + 10474464 y^{10} + 90679680 y^{11} \\
& \quad + 544536720 y^{12} + 2010441888 y^{13} + 3446493984 y^{14}.
\end{align*}
\item $(17,5^{17},9)$, 
$(\omega 0010111001110100)$,
\begin{align*}
W(1,y) &= 1 + 97240 y^9 + 1633632 y^{10} + 24504480 y^{11} + 296652720 y^{12} \\
& \quad + 2733620400 y^{13} + 18749403360 y^{14} + 89994568992 y^{15} \\
& \quad + 269984494620 y^{16} + 381154477680 y^{17}.
\end{align*}
\item $(18,5^{18},10)$, $(\omega 12134242124243121)$,
\begin{align*}
W(1,y) &= 1 + 1050192 y^{10} + 11456640 y^{11} + 180442080 y^{12} \\
& \quad + 1964813760 y^{13} + 16877613600 y^{14} + 107991522432 y^{15} \\
& \quad + 485972877960 y^{16} + 1372155934320 y^{17} + 1829541554640 y^{18}.
\end{align*}
\end{itemize}

As mentioned in the introduction, stabilizer codes can be defined over any Abelian group, not
only finite fields. For comparison, we also generated circulant codes over $\mathbb{Z}_4^2$.
As expected, the minimum distance of these codes are much worse than for codes over $\mathbb{F}_{16}$.
We found a $(7,4^7,4)$-code over $\mathbb{Z}_4^2$, but for all other lengths up to 16, the best 
minimum distance was equal to the best minimum distance of codes over $\mathbb{F}_4$ of the same length.

Gulliver and Kim~\cite{gulliverkim} performed a computer search of circulant self-dual
additive codes over $\mathbb{F}_4$ of length up to 30. Their search was not restricted to 
graph codes, so our search space is a subset of theirs. It is interesting to note that for every length,
the highest minimum distance found was the same in both searches. 
This suggests that the circulant graph code construction can produce codes as strong as the
more general circulant code construction.
Besides a smaller search space, the special form of the generator matrix of a 
graph code makes it easier to find 
the minimum distance, since any codeword obtained as a linear combination of $i$ rows 
of the generator matrix must have weight at least $i$. If, for example, we want to determine whether
a code has minimum distance at least $d$, we only need to consider combinations of $d$ or 
fewer rows of its generator matrix.

\begin{figure}
 \centering
 \subfloat[{$(6,2^6,4)$}]
 {\hspace{3pt}\includegraphics[width=.43\linewidth]{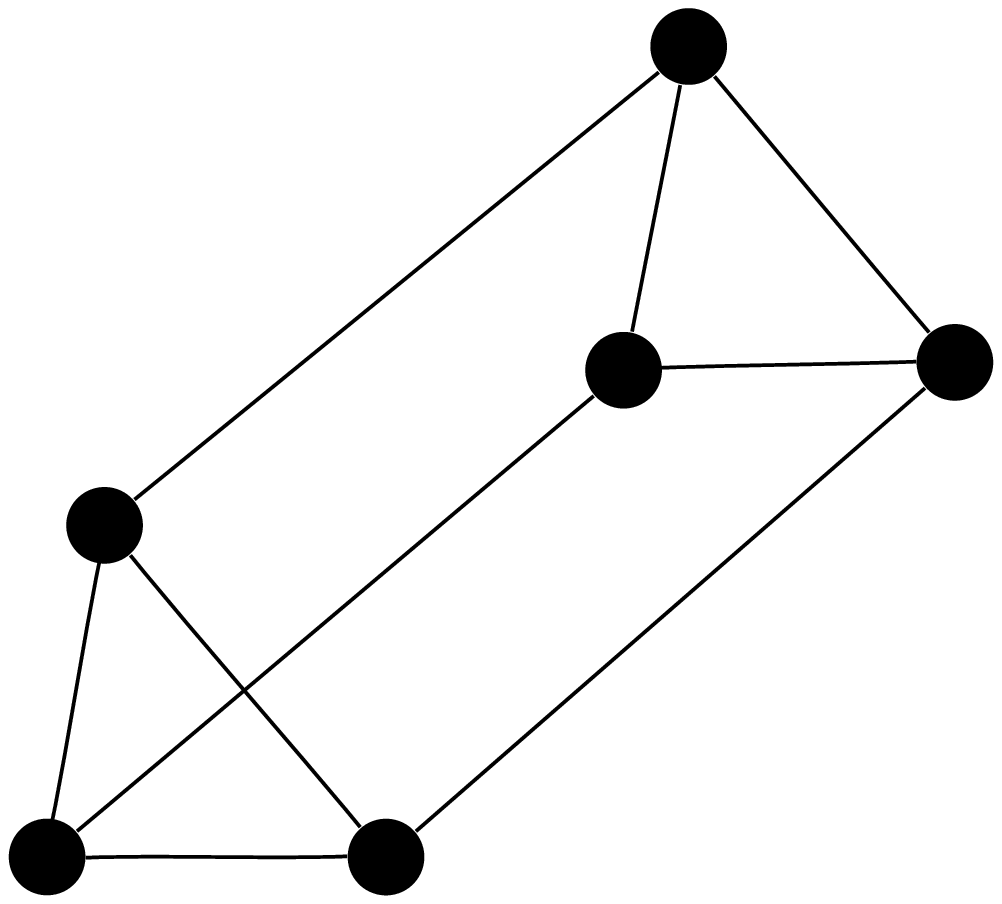}\hspace{3pt}\label{2cl3cl}}
 \subfloat[{$(12,2^{12},6)$}]
 {\hspace{3pt}\includegraphics[width=.43\linewidth]{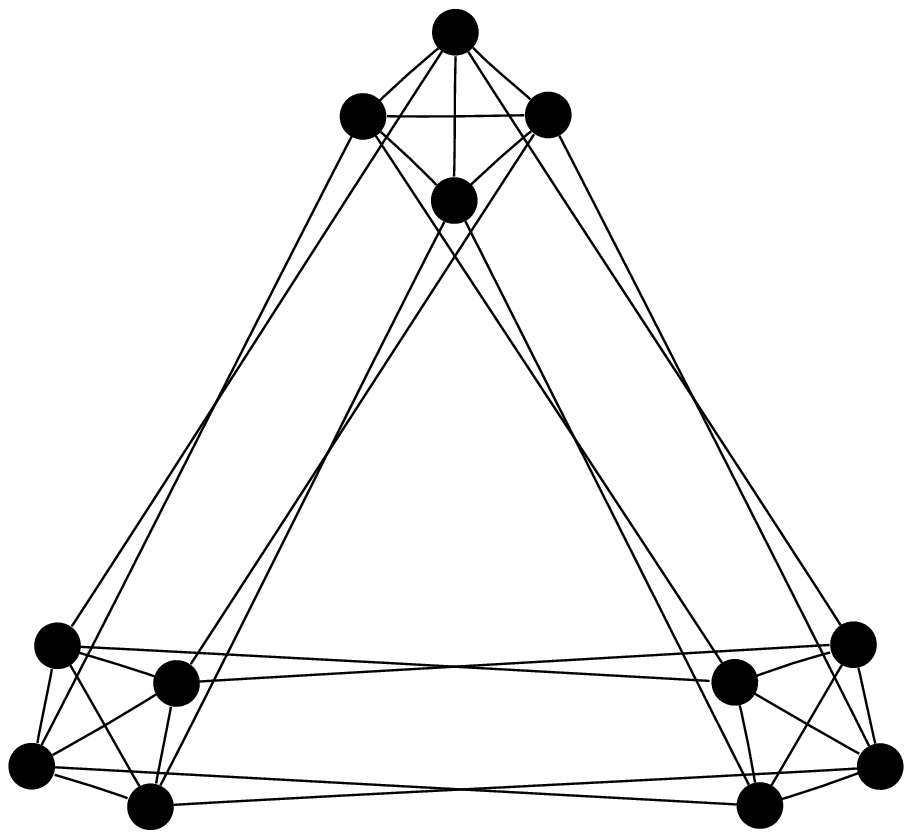}\hspace{3pt}\label{3cl4cl}}
 \subfloat[{$(20,2^{20},8)$}]
 {\hspace{3pt}\includegraphics[width=.43\linewidth]{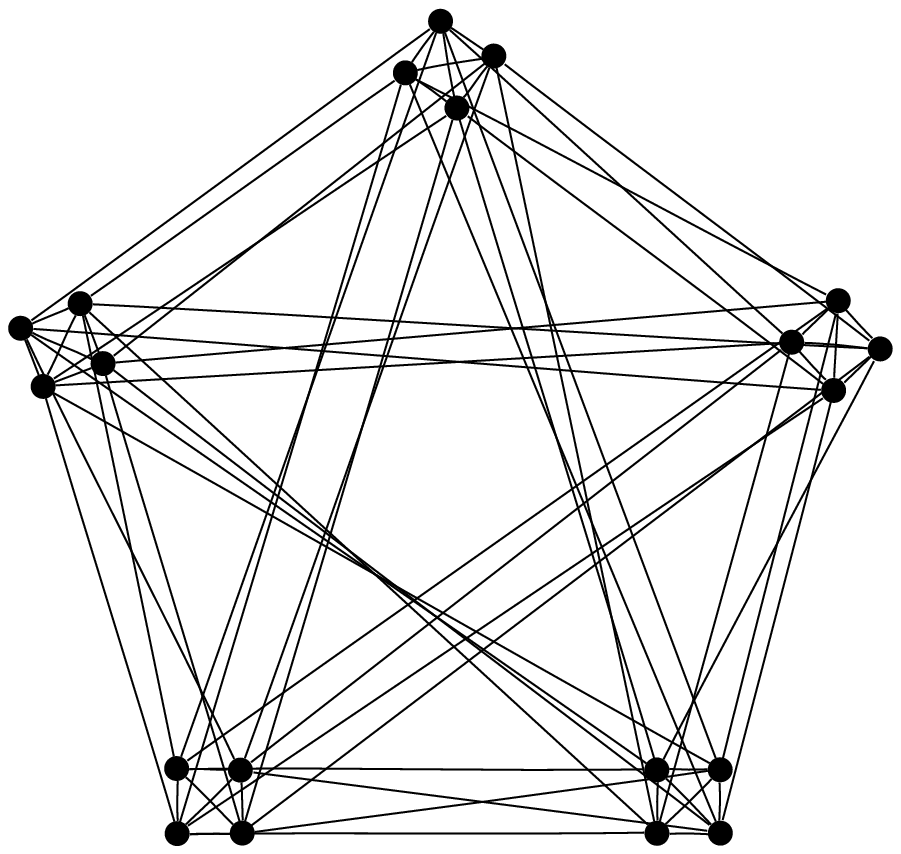}\hspace{3pt}\label{5cl4cl}}
 \subfloat[{$(25,2^{25},8)$}]
 {\hspace{3pt}\includegraphics[width=.43\linewidth]{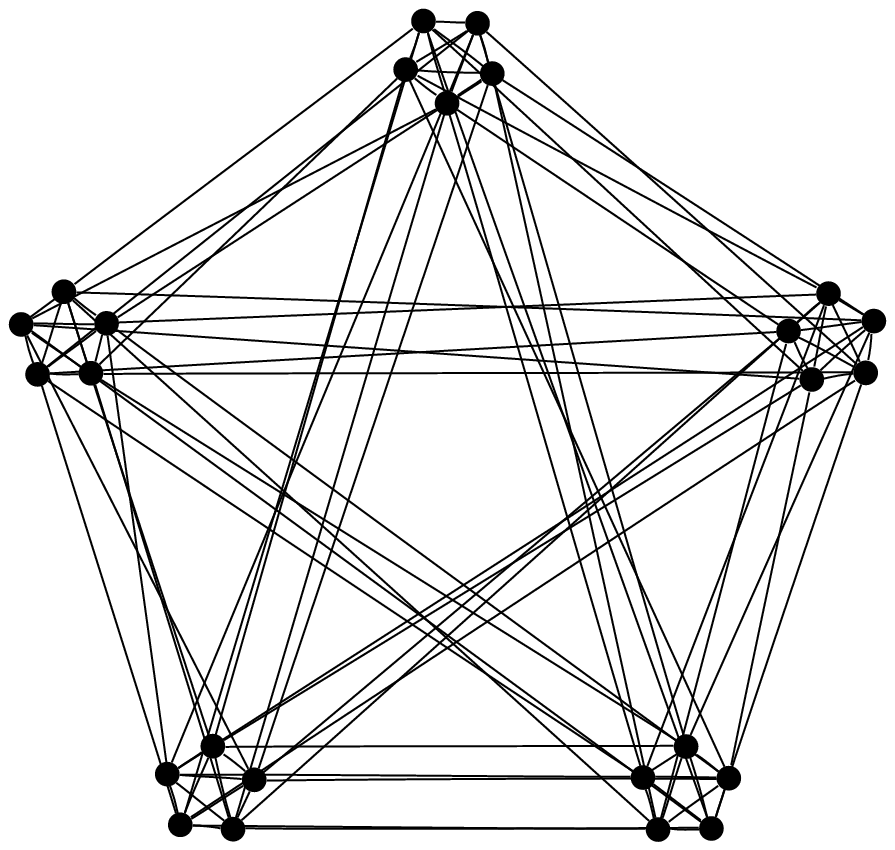}\hspace{3pt}\label{5cl5cl}}
 \caption{Examples of Nested Clique Graphs Corresponding to Codes over $\mathbb{F}_4$}\label{graphs}
\end{figure}

Circulant graphs must be \emph{regular}, i.e., all vertices must have the same number of 
neighbours. We have previously discovered~\cite{mscthesis,setapaper} that many strong circulant
self-dual additive codes over $\mathbb{F}_4$ can be represented as highly structured \emph{nested clique
graphs}. Some of these graphs are shown in Fig.~\ref{graphs}. For instance, Fig.~\ref{3cl4cl} shows
a graph representation of the $(12,2^{12},6)$ ``Dodecacode'' consisting of three 4-cliques. The remaining
edges form a \emph{Hamiltonian cycle}, i.e., a cycle that visits every 
vertex of the graph exactly once.
Notice that all graphs shown in Fig.~\ref{graphs} have \emph{minimum regular vertex degree}, i.e.,
each vertex has $d-1$ neighbours, where $d$ is the minimum distance of the corresponding code.

\begin{figure}
 \centering
 \includegraphics[height=0.4\linewidth]{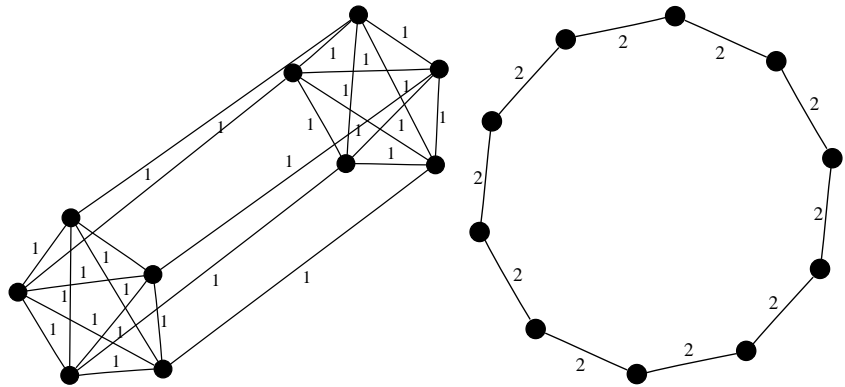}
 \caption{Two Graphs Whose Sum Corresponds to the $(10,3^{10},6)$ Code}\label{3graph}
\end{figure}

We have discovered some new highly structured weighted graph representations of self-dual additive
codes over $\mathbb{F}_{9}$ and $\mathbb{F}_{16}$. Fig.~\ref{3graph} shows two interconnected 5-cliques
where all edges have weight one, and a 10-cycle where all edges have weight two. The sum of these two graphs, such that
no edges overlap, corresponds to the $(10,3^{10},6)$ code. 
Up to isomorphism, there is only one way to add a Hamiltonian cycle of weight two
edges to the double 5-clique, since there cannot be both weight one and weight two edges between the same
pair of vertices.
The first row of a circulant generator matrix corresponding to this graph is $(\omega012111210)$.

\begin{figure}
 \centering
 \includegraphics[height=0.3\linewidth]{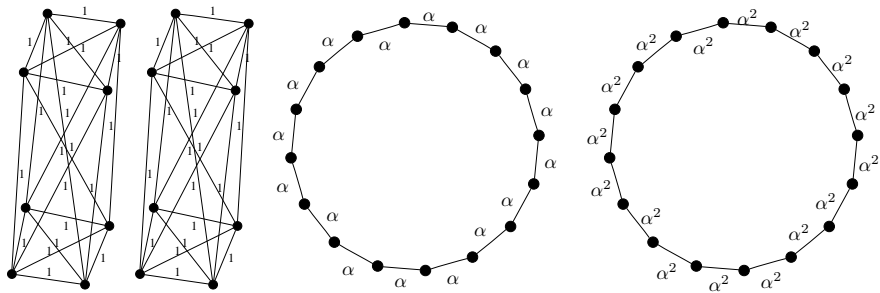}
 \caption{Three Graphs Whose Sum Corresponds to the $(16,4^{16},8)$ Code}\label{4graph}
\end{figure}

As a second example, Fig.~\ref{4graph} shows two pairs of 4-cliques, each of
which is connected by a length 8 cycle, and two 16-cycles where all edges have 
weight~$\alpha$ and $\alpha^2$, respectively, where $\alpha = \omega^5$ is a primitive element of
$\mathbb{F}_4$.
The $(16,4^{16},8)$ code generated by $(\omega0\alpha^21\alpha100010001\alpha1\alpha^2)$ 
corresponds to a sum of these three graphs.
 
Note that the vertices of the graphs corresponding to circulant $(10,3^{10},6)$ and $(16,4^{16},8)$ 
graph codes have degree higher than $d-1$.
We have tried to obtain similar graph representations for other codes in Table~\ref{tab:circulant},
but without success. Many of the circulant graph codes have vertex degree much higher than $d-1$, 
for instance the $(14,5^{14},8)$ code generated by $(\omega1221202021221)$,
and the $(18,5^{18},10)$ code generated by $(\omega12134242124243121)$.

\section*{Acknowledgements}
The author would like to thank Matthew G. Parker for helpful discussions and comments.
Also thanks to Markus Grassl for helpful comments.

\end{document}